%% file: gizis.tex
\documentclass[preprint]{aastex}

\shortauthors{Gizis et al.}
\shorttitle{HST Binaries}

\begin{document}

\title{Hubble Space Telescope Observations of Binary Very-Low-Mass Stars and Brown
Dwarfs}

\author{John E. Gizis\altaffilmark{1}}
\affil{Department of Physics and Astronomy, University of Delaware,
Newark, DE 19716 \email{gizis@udel.edu}}

\author{I. Neill Reid}
\affil{Space Telescope Science Institute,
3700 San Marin Drive, Baltimore, MD 21218 }

\author{Gillian R. Knapp}
\affil{Department of Astrophysical Sciences, Princeton University, Princeton, NJ 08544-1001}

\author{James Liebert}
\affil{Steward Observatory, University of Arizona, Tucson, AZ 85721}

\author{J. Davy Kirkpatrick}
\affil{Infrared Processing and Analysis Center, 100-22, California Institute of Technology, Pasadena, CA 91125}

\author{David W. Koerner}
\affil{Department of Physics and Astronomy, Northern Arizona University,
NAU Box 6010, Building 19, Flagstaff, AZ  86011-6010}

\author{Adam J. Burgasser\altaffilmark{2}}
\affil{UCLA Division of Astronomy \& Astrophysics, 8965
Math Science Bldg., 405 Hilgard Ave., Los Angeles, CA, 90095-1562}

\altaffiltext{1}{Visiting Astronomer, Kitt Peak National Observatory,
National Optical Astronomy Observatories, which is operated by
the Association of Universities for Research in Astronomy, Inc.
(AURA) under cooperative agreement with the National Science Foundation.}
\altaffiltext{2}{Hubble Fellow}

\begin{abstract}
We present analysis of Hubble Space Telescope images of 82 nearby field
late-M and L dwarfs. We resolve 13 of these systems into
double M/L dwarf systems and identify an additional possible binary. 
Combined with previous observations of
20 L dwarfs, we derive an observed binary fraction for ultracool
dwarfs of $17^{+4}_{-3}\%$, where the statistics 
included systems with separations in
the range 1.6-16 A.U.    We argue that
accounting for biases and incompleteness leads to an estimated 
binary fraction $15\pm5$\% in the range 1.6-16 A.U. No systems wider
than 16 A.U. are seen, implying that the wide companion frequency
is less than 1.7\%; the distribution of orbital separation is peaked at $\sim 2-4$ A.U. and differs greatly from the G dwarf binary distribution.
Indirect evidence suggests that the binary fraction is $\sim 5\pm3\%$ for
separations less than 1.6 A.U.
We find no evidence for differences in the binary
fraction between stellar late-M and L dwarfs and substellar 
L dwarfs.  We note, however, that the widest ($>10$ A.U.) systems in 
our sample are all of earlier (M8-L0) spectral type; a larger
sample is needed determine if this is a real effect. 
One system with a spectral type of L7 has a secondary that is
fainter in the HST F814W filter but brighter in F1042M; we 
argue that this secondary is an early-T dwarf.
\end{abstract}


\section{Introduction}

The Deep Near-Infrared Survey \citep[DENIS]{denis,delfosse}, 
Two Micron All-Sky Survey
\citep[2MASS]{mike,2mass},
and Sloan Digital Sky Survey \citep[SDSS]{sdss} have enabled the discovery
of numerous cool dwarfs and led to the definition of new spectral types
L \citep{k99,mldwarf} and T \citep{2masst,sdsst}. 
With effective temperatures lower than
$\sim2000$K and $\sim1300$K respectively, these objects span the
range between the lowest mass stars and sub-stellar mass brown dwarfs.
The latest M dwarfs (M8/M9) and the earlier-type  L dwarfs represent
a mix of brown dwarfs which are still cooling and long-lived hydrogen-burning
stars which have reached a stable state \citep{reidmf}. Theoretical
models indicate that the overwhelming majority of L dwarfs
with temperature below $\sim1750$K (spectral types $\approx$L5 and later;
Kirkpatrick et al. 1999,  Gizis et al. 2000),
and all T dwarfs, are expected to be brown dwarfs.

The large samples now available of these objects
allow investigation of their statistical characteristics. One
goal of such studies is the establishment of empirical constraints
on the formation mechanism(s) of low mass dwarfs, particularly
searching for potential differences in the properties of the
lowest mass stars and brown dwarfs. It is clear that
the properties of binary brown dwarfs are important in
this context. A comprehensive
theory of star formation must account not only for
the stellar/substellar initial mass function (IMF), but also the
frequency and orbital distributions of binary systems.
Although
brown dwarfs are extremely rare as close companions to GFK stars
\citep{hippbd,mb00} a number
of studies have detected double brown dwarf systems
\citep{martinbinary,koerner,reidhst,close1,adamhst}, and even 
doubles that orbit more massive stars \citep{gl569b,potter}.
\footnote{After the initial submission of this paper, \citet{close3} 
reported the independent discovery that 2M1127+74AB and 2M1311+80AB
are doubles.}

The present paper presents analysis of the largest sample of
high-resolution images of low mass dwarfs assembled so far.
We have used the Planetary Camera of WFPC2 on the
Hubble Space Telescope WFPC2 to image a sample of 82 late-type M and
L dwarfs. Details of those observations are presented in Section~\ref{data}.
Comments on individual systems are given in Section~\ref{systems}.
We have combined our present observations with data for
twenty L dwarfs analyzed by \citet{reidhst}, and use those data to
set constraints on the binary fraction in low mass dwarfs. That
analysis is presented in Section~\ref{binfrac}, and our conclusions
summarized in the final section.

\section{Data\label{data}}

\begin{figure}
\epsscale{0.8}
\plotone{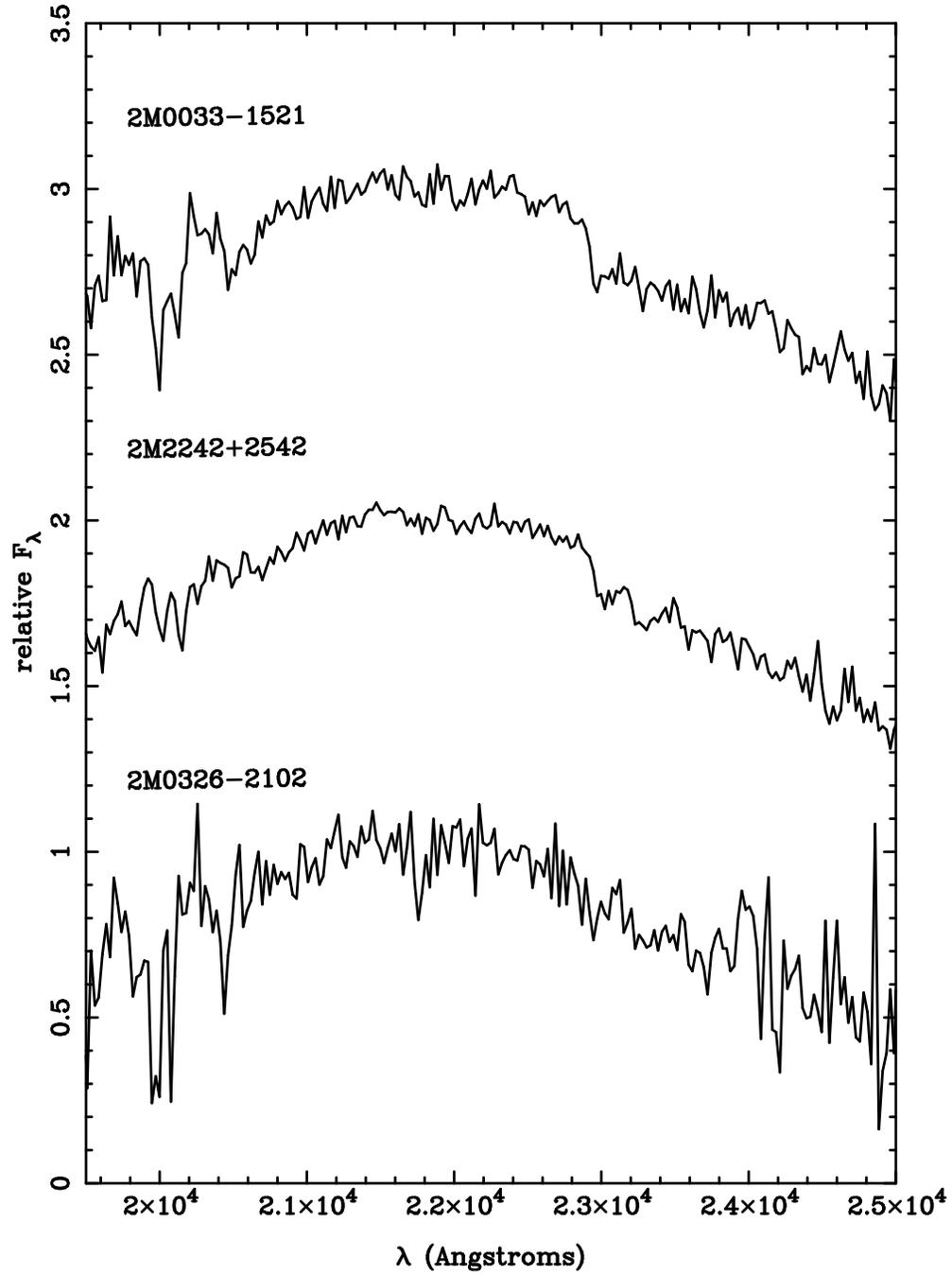}
\caption{KPNO 4-meter CRSP spectra of three new L dwarfs.  The observations
are described in \citet{andreas2}.
\label{fig-lnew}}
\end{figure}

Targets for Snapshot WFPC2 imaging with HST were chosen from the list of
2MASS-selected
late-M dwarfs \citep{gizis00} and  L dwarfs
\citep{k99,k00}, and from late-M and L dwarfs identified from
the initial regions covered by SDSS \citep{fan,sdssl2,hawley}.
A few 2MASS and SDSS objects included have not yet been described in
publications.   Near-infrared discovery spectra for 2M0033-1521,
2M0326-2102 and 2M2242+2542 are shown
in Figure~\ref{fig-lnew}.
These data were obtained using the infrared Cryogenic Spectrometer (CRSP)
on the Kitt Peak 4-meter telescope during observing runs
in February and November 1999.  The February data were obtained
using Grating 2.  The setup provided L-band spectra from
K-band spectra from 1.85 to 2.55 $\mu$m.  In November,
we switched to the slightly higher resolution Grating 4, which
yielded spectra over the range 1.90 to 2.55 $\mu$m at K-band
or 1.05 to 3.75 $\mu$m at J-band.   The resolution, as
measured by the FWHM of arc lines, was 55\AA~ and 110 \AA~
at J and K respectively for Grating 4.  Data from this run are
discussed in \citet{andreas2}. 
Observations were obtained in the standard way described in
the CRSP User's Manual \citep{crsp}.  Each object was stepped
along the slit to obtain observations at five positions.
When necessary, this cycle was repeated to obtain higher
signal-to-noise.  Typical individual exposures were
20 seconds at K-band.  Spectra were extracted using IRAF.  
Both the HST photometry described below and 
comparison of these near-IR spectra
to 25 optically classified L dwarfs observed with the same setup
indicates that they are L dwarfs; however, optical spectra are needed
to obtain accurate spectral types.  

Each object on the HST program was observed at least once
using the F814W and F1042M filters.  For the brighter targets, a
second F814W observation was possible in the timeframe allowed by the
Snapshot visit.  Exposure times in F814W ranged from 40 to 400 seconds
while exposure times in F1042M were always 500 seconds.
A total of 82 dwarfs were observed (Table 1).  In general, 
\citet{gr95} found that we are able to resolve 
systems with magnitude differences of 0, 1, 3 and 5 
at separations of 0.09, 0.14, 0.23, and 0.31 arcseconds
respectively using WFPC2.  It should be noted, however, than in the
case of the fainter L dwarfs, necessarily observed with lower
signal-to-noise, that only companions within 3 magnitudes
of the primary are detectable.  We searched only the PC chip
for companions.  Coverage for wider companions is provided by 
the 2MASS data,
which would have detected any companions ($K_s < 14.5$) 
wider than a few arcseconds.  

Aperture photometry was measured and zero-pointed
for each target according to the
precepts of \citet{hstphot1}.  However, since we usually
have only single frames in
each filter, we found that the large aperture
magnitudes were often contaminated by cosmic rays.  Furthermore,
the aperture magnitudes cannot be used for close doubles.
DAOPHOT/IRAF was therefore used to measure magnitudes by
fitting a PSF model, which was built using images of bright,
single stars from this program.  The PSF-subtracted images
were an effective way of verifying that no secondaries were
missed in the apparently single stars.  The normalization between
the DAOPHOT magnitudes and the aperture magnitudes was determined
by taking the median value for high signal-to-noise, isolated stars.
Finally, we corrected for charge transfer efficiency (CTE) losses
using \citet{dolphin}'s Y-CTE corrections and X-CTE corrections.
The resulting magnitudes for each system are given in
Table~\ref{table-phot}.  This table includes 2MASS Version-2
Working Database (corresponding to the Second Incremental Data Release 
system) JHK$_S$ magnitudes for all systems.  None of the binaries were
resolved by 2MASS, so the magnitudes listed reflect the combined
photometry. Table~\ref{table-phot} lists composite WFPC2
photometry for those systems. Magnitudes and separations for the
individual components in the resolved systems are given in
Table~\ref{table-binaries}.   Z-band photometry of
2M0335+2342, 2M1711+2232, 2M2208+2921, and 2M2306-0502 
was not possible due to cosmic ray hits.
It is possible that low-level cosmic ray hits might also affect 
other targets, especially the Z band.

Two of the systems (2M0345+2540,2M2224-0158) 
in our sample have trigonometric parallax measurements
\citep{dahn}.
For the remainder, we have estimated photometric parallaxes.
Unfortunately, the $I-Z$ WFPC2-based colors are
not of high enough accuracy to directly estimate photometric distances
for each component. Our distances are therefore based on
combining the observed I magnitude for the primary,
the \citet{reidhst} $I_C-I_{814}$ relation, and
the \citet{dahn} $I_C-J$,$M_J$ relation.\footnote{We have adopted
this procedure because our new sample adds only one object
with an I$_C$ measure, and because many of the parallax stars lack
WFPC2 measurements.}  We assume the
system $I-J$ color corresponds to the primary's $I-J$ color in order
to use the \citet{dahn} relation; we tested
differing assumptions such as using the observed system spectral type
and got consistent answers.
The transformation from separation in
arcseconds to A.U. is uncertain by $\sim 30\%$; trigonometric
parallaxes are needed.  These errors have no significance for the
interpretation of the data.  

\begin{figure}
\epsscale{0.8}
\plotone{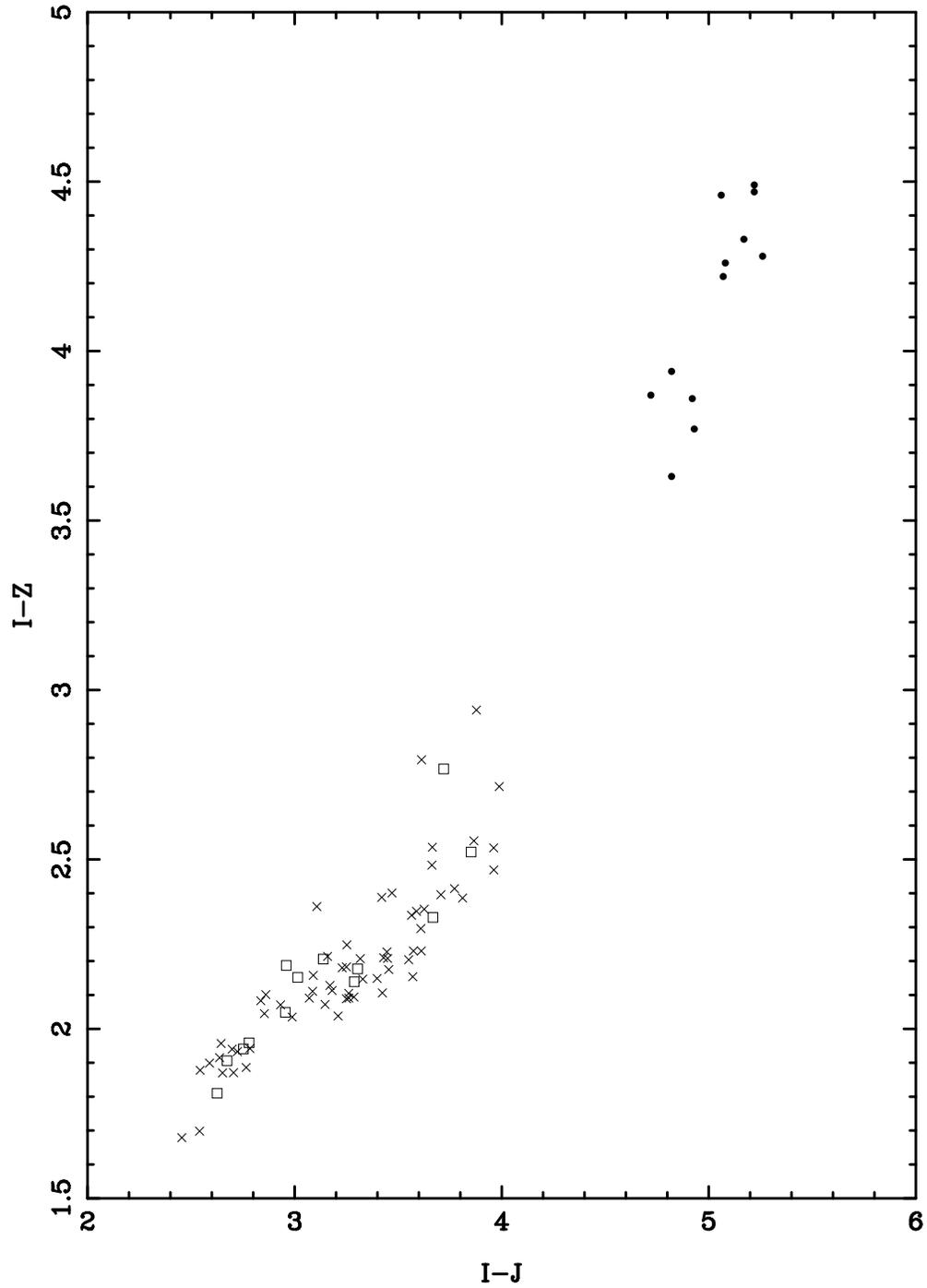}
\caption{Color-color diagram using HST (I,Z) and 2MASS (J) photometry.
Apparently single dwarfs are shown as crosses and doubles as open squares.
T dwarfs from \citet{adamhst} are shown as solid points.  
\label{fig-colors1}}
\end{figure}

\begin{figure}
\epsscale{0.8}
\plotone{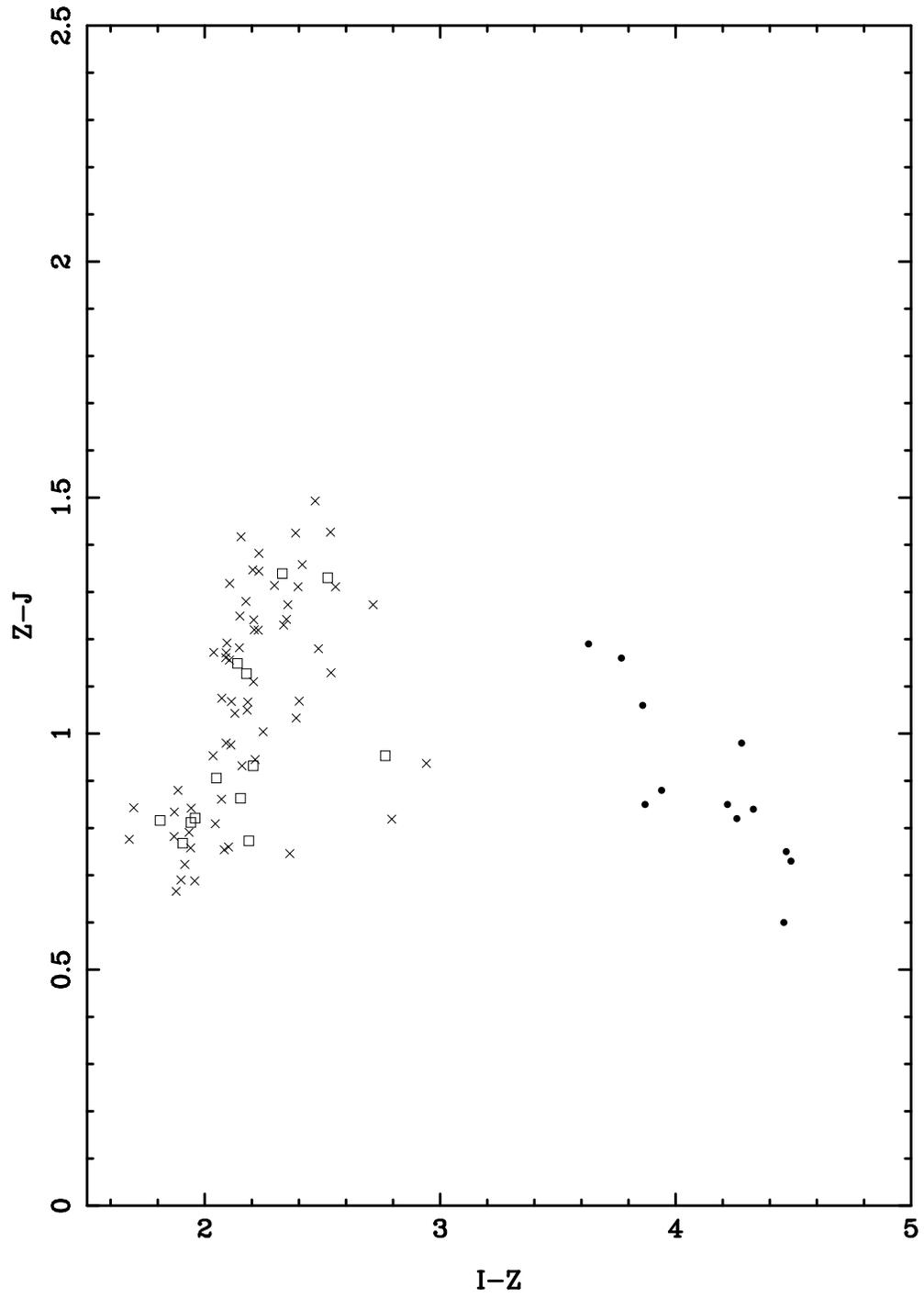}
\caption{Color-color diagram using HST (I,Z) and 2MASS (J) photometry.
Symbols as in Figure~\ref{fig-colors1}
\label{fig-colors2}}
\end{figure}

Color-color diagrams combining the HST and 2MASS photometry are
shown in Figures~\ref{fig-colors1} and ~\ref{fig-colors2}.
T dwarfs with WFPC2 photometry by \citet{adamhst} are also shown. 
The gap between the latest L dwarfs of our program and the
T5 and later dwarfs is obvious and is due to the lack of WFPC2 
observations of early-T dwarfs.  Our reddest object in I-Z is
the latest L dwarf, the L8 2M0328+2302.  

\section{Individual Systems\label{systems}}

{\bf 2M1239+5515:} This system shows lithium absorption according to
\citet{k00}, the only resolved binary in this study to do so.  As
such, both components of the system are below the lithium-burning
limit of $0.055-0.060 M_\odot$ \citep{chabrier,burrev}.

{\bf 2M1728+3948:}  This system was classified L7 by \citet{k00}.
Remarkably, component A is brighter by 0.3 magnitudes at I but
{\it fainter} by 0.3 magnitudes at Z than component B.  Component B,
with I-Z color of 3.1, is redder than the latest-type L dwarfs
(Figures~\ref{fig-colors1} and~\ref{fig-colors2}), 
suggesting that it lies in the L/T transition r\'egime.  
This behaviour invites comparison with observations of  late-L
and early-T dwarfs at 1.25$\mu$m, where the latter dwarfs
are observed to be brighter by up to 1 magnitude \citep{dahn}.
This has been interpreted as evidence for cloud disruption at
the atmospheric temperatures spanned by the
L/T transition \citep{cloud}. Our observations provide the
first evidence that the effect extends down to 1 micron.
The primary in this system
is cool enough that both components are brown dwarfs,
with at least the primary above the lithium-burning limit.

{\bf 2M1017+1308, 2M1430+2915}: These two objects were
identified as L dwarfs by J. Wilson (2002, priv. comm.) using
near-IR spectra.  The HST colors confirm the cool dwarf classification.
Given the spectral types, all of the components could
be either stars or brown dwarfs.

{\bf 2M1449+2355}: The limits on lithium for this distant
L0 dwarf \citep{k00} system do not rule out its presence.
Both components may be stars or brown dwarfs.   At 13 A.U., this is the
most widely-separated L dwarf double currently known.

{\bf 2M1600+1708}: LRIS observations of this L1.5 dwarf \citep{k00} system
show no evidence for lithium absorption, indicating $M > 0.055 -0.060 M_\odot$.
The HST image is better modelled by two
PSFs separated by 0.056 arcseconds than by a single PSF; we regard this
is only a {\it candidate} double.  Additional
observations are required to demonstrate that this truly
is a binary system.

{\bf 2M2101+1756}: This L7.5 dwarf \citep{k00} system is
late enough that both components must be brown dwarfs.
The secondary is probably an L8 or slightly later dwarf.

{\bf 2M1426+1557,2M2140+1625,2M2206-2047,2M2331-0406:}  These late-M dwarf
systems have primaries that are more likely to be stars
than brown dwarfs; if any of these systems are young then the
primary may be a brown dwarf.  
All  four systems were discovered by
\citet{close1,close2} using Gemini Adaptive Optics J, H and
K-band imaging, and to whose discussion of individual systems
we refer the reader.  The $IZ$ photometry presented here is
consistent with the spectral types expected from the JHK photometry.
  2M2206-2047 has a low tangential  velocity (10 km/s), which
is consistent with youth; since many old stars have small velocities, 
it does not prove youth.  \citet{reidhires} have published a
high resolution spectrum  of 2M2206-2047 which shows rapid rotation and
no lithium; thus, the possibility that the primary is a high-mass brown dwarf 
remains viable, but the mass must be above the lithium-burning limit.  

{\bf 2M1127+7411,2M1311+8032,2M2147+1431:}  These late-M dwarf
systems have primaries that are more likely to be stars
than brown dwarfs; if any of these systems are young then the
primary may be a brown dwarf. Using our distance estimate and the
\citet{gizis00} proper motion, the tangential velocity of
2M1127+7411 is only 6 km/s; this suggests youth but is not 
definitive.  Our measurements of 2M1127+74AB and
2M1311+80AB are consistent with those reported by \citet{close3}.

{\bf 2MASSW0335+2342}:  This M8.5 dwarf shows lithium absorption
and therefore is young brown dwarf \citep{reidhires}.  Our observations
show no evidence for binarity.

\section{The Binary Fraction\label{binfrac}}

The most notable characteristic of the data is the lack of wide
systems.  Even for our faintest, most distant targets, we would
have resolved companions with separations $>20$ A.U., yet none are
seen.  Combining our sample with \citet{reidhst}, we find that
0 out of 102 late-M and L dwarfs have such companions with $\Delta I<3$
(and in many cases, $\Delta I<5$).  The 1-sigma upper limit
on the $>20$ A.U. companion frequency is then 1.7\%.  

\begin{figure}
\epsscale{0.8}
\plotone{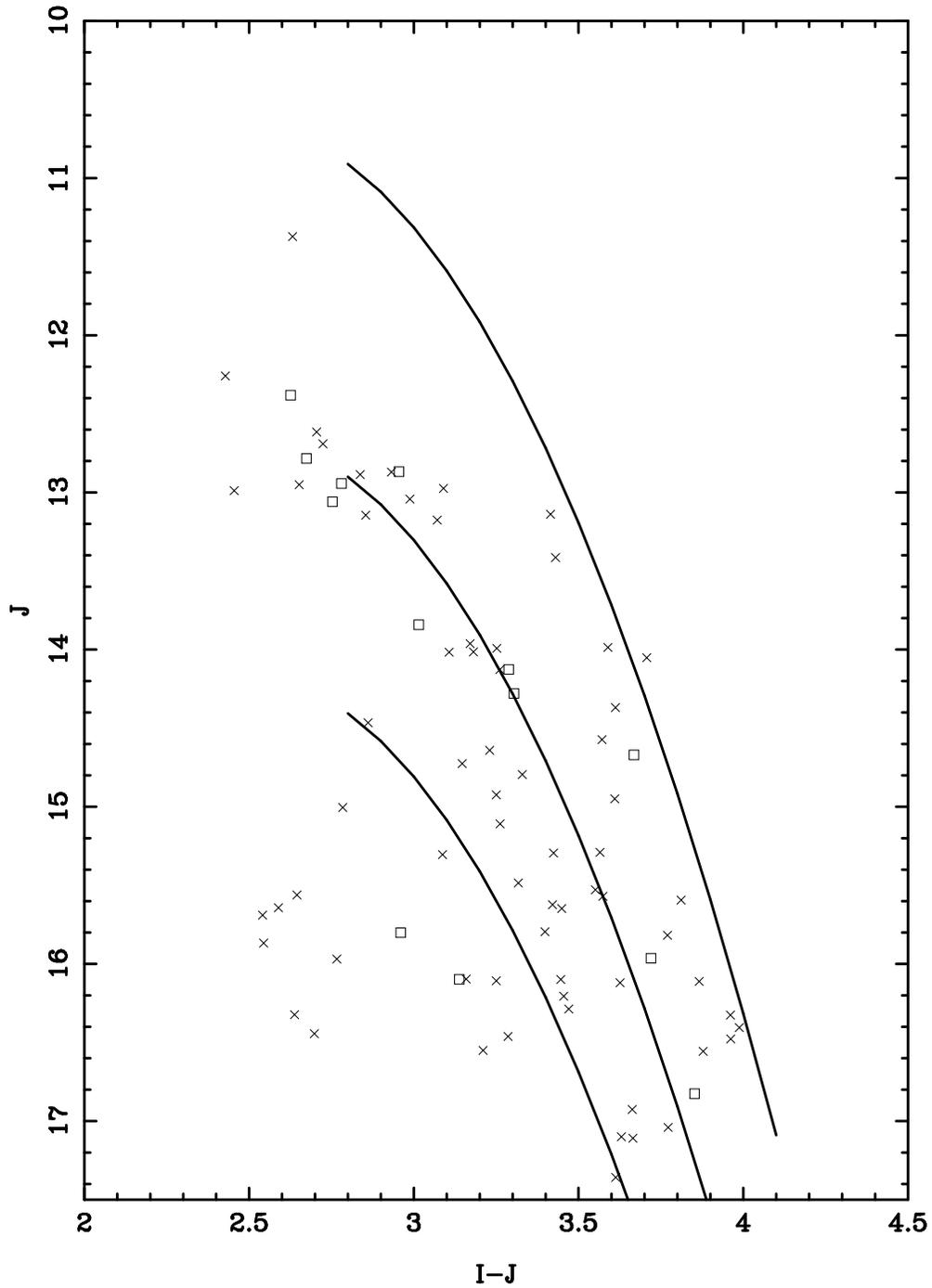}
\caption{Color-magnitude diagram for the sample without corrections for
binarity.  The apparently single systems appear as crosses and the
resolved doubles as open squares.  The solid curves plot the \citet{dahn}
cool dwarf sequence at distances of 10, 25, and 50 parsecs.  
\label{fig-distance}}
\end{figure}

\begin{figure}
\epsscale{0.8}
\plotone{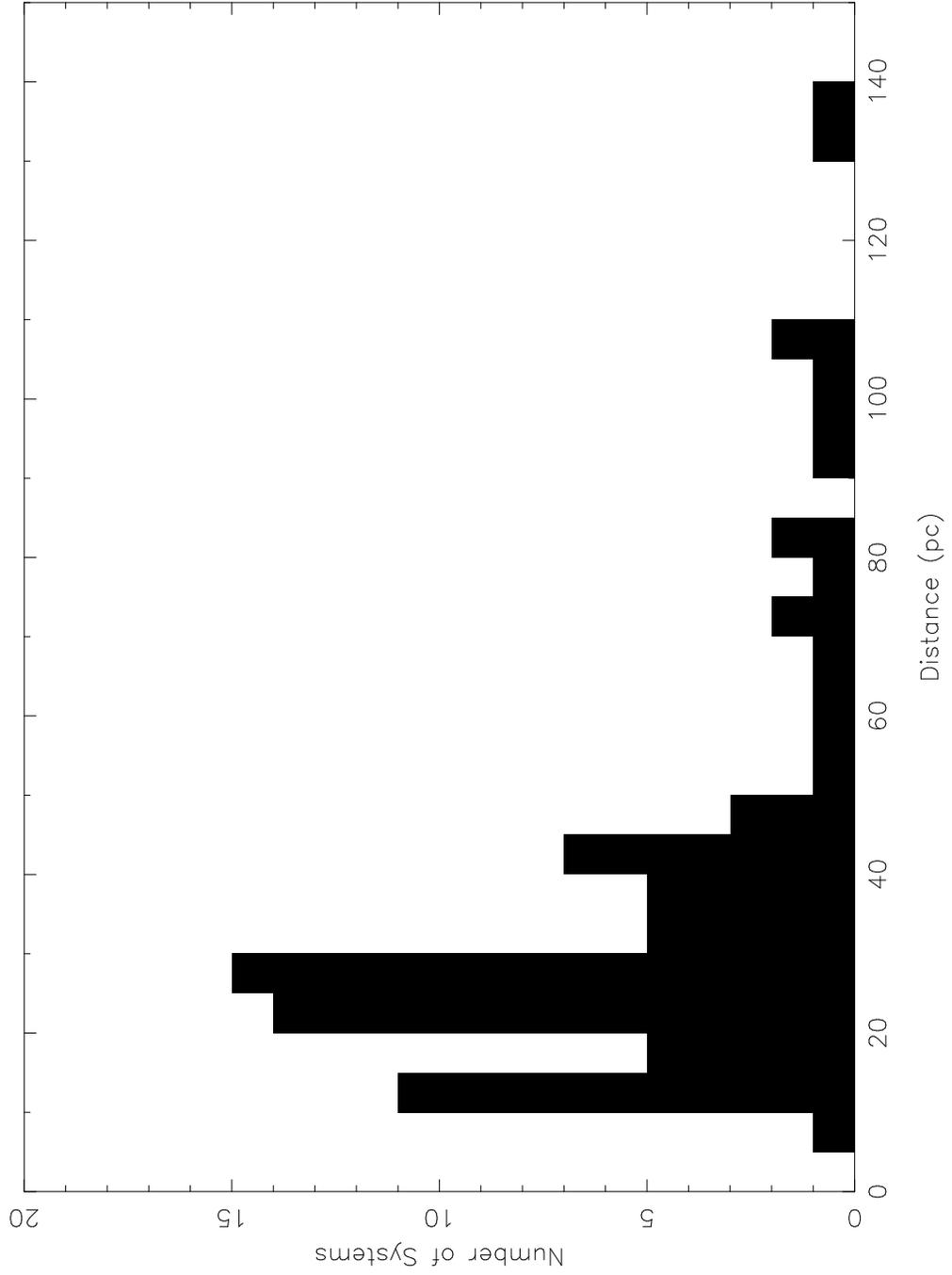}
\caption{Histogram of estimated photometric parallaxes.  
\label{fig-dhist}}
\end{figure}

\begin{figure}
\epsscale{0.8}
\plotone{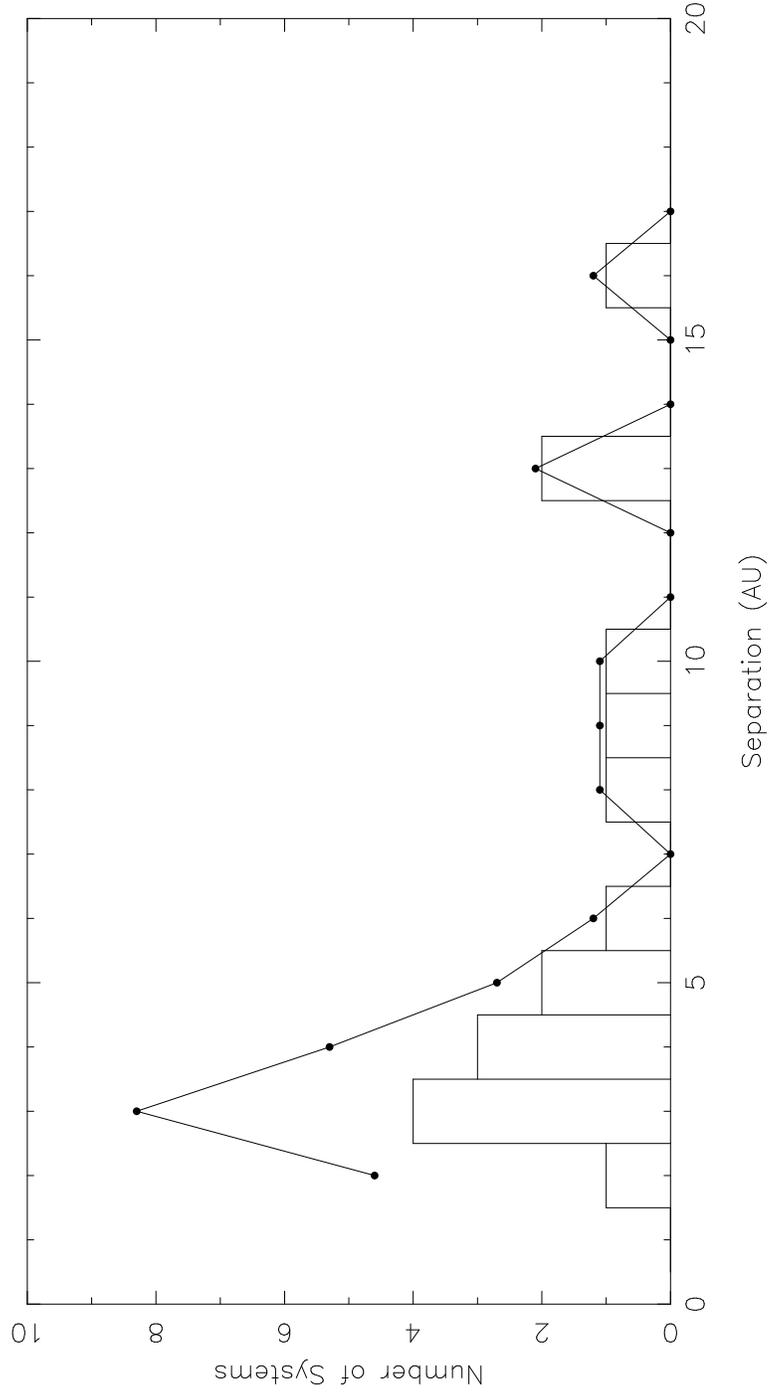}
\caption{Histogram of observed orbital separations.  The histogram shows
the count of resolved systems, while the solid points with connecting lines
shows the distribution after correction for the fact that the closest systems
cannot be resolved in all cases.  The distibution is peaked, with companions
more common in the range 2-5 A.U. in the range 6-10 A.U.  
\label{fig-sephist}}
\end{figure}

Thirteen of the 82 systems in the present sample are resolved as binaries
with component separations exceeding our detection thresholds, all in the 
range 3-16 A.U. 
We have combined 
these data with the sample of twenty L dwarfs from \citet{reidhst}.
Those dwarfs were also identified by color selection from the 2MASS
database, and four are clearly resolved as binary systems with $\Delta > 0.06$
arcsec. 
This total observed binary fraction of 17$^{+4}_{-3}$\%, where the
uncertainties are calculated according to \citet{adamhst},  must be corrected 
to obtain a true binary fraction.   First,
some of the systems are more distant than others, making the tighter
systems unresolvable.  The importance of this effect is evident
in Figure~\ref{fig-distance}, which shows the magnitudes and colors
of the target systems.  Four of the ten systems with $I-K<4$ and
$K<12$
are resolved as doubles, yet none of the seven systems with
$I-K<4$ and $K>13.5$ are resolved.  The latter are typically
four times further away, making most double systems unresolvable.
Figure~\ref{fig-dhist} plots the histogram of the distance estimates.
Overall, excluding the 16 systems in Figure~\ref{fig-dhist}
with photometric distances greater than 50 parsecs, {\it including
the ``wide'' binary 2M1449+2355 and candidate binary 2M1600+1708}, 
the observed binary fraction is increased to $19\pm4$\%.  
Nevertheless, we may still fail to resolve many of the 
systems in the 1.6-10 A.U. range.  This effect biases both the estimate
of the binary frequency and the derived 
binary separation distribution.  The observed binary separation 
distribution is plotted in Figure~\ref{fig-sephist}.  In order to
account for the systems which cannot be resolved, we also weight each
system by the number of target systems in which we could have resolved 
it.  For example, the binary 2M1017+1308 could have been resolved if
it had been as close as 0.10 arcseconds; given our distance estimates
it might have been resolved in only 57 of the total 102 targets searched.  
The resulting corrected distribution is also shown in Figure~\ref{fig-sephist}.
It is evident that binaries in the range $\sim2-5$ A.U. are more common
than the wider systems.  

A second important effect is that a
luminosity-selected sample such as this one will be afflicted
by systematic biases.
Two conflicting biases exist \citep{reid91}.  First, near-equal luminosity
binaries can be seen to greater distance and therefore are over-represented
in the sample; if these extra doubles are
resolved they will {\it increase} the observed binary fraction over
the true one.  Second, if these extra doubles are not resolved, they will
{\it decrease} the observed binary fraction since they are counted as
single stars.  In the case of late-M and L dwarfs, where there are no
observed systems with separations $\gtrsim 15$ A.U., the binaries may
usefully
be considered in two groups:  Those with separations in the range
$1.6-15$ A.U. and those with separation $<1.6$ A.U.  Our WFPC2 observations
can resolve only the former group (with the innermost
limit worse for the more distant objects), but through the second bias effect
the latter group can affect our measurements.

The situation can be illustrated by a simple Monte Carlo
model.  For illustrative purposes only, we consider L dwarfs
to be uniformly distributed in space out to a distance of 50 parsecs
to a limiting magnitude of $K_s=14.0$.  In the present
schematic model, we assign the same absolute magnitude, 
$M_K=11.5$, to each primary.  
Secondary companions are added to
a fraction $f_1$ of the L dwarfs at separations $<1$ A.U. (unresolvable)
and a fraction $f_2$ at $1-10$A.U. (potentially resolvable).  The
K-band flux ratios are distributed uniformly in the range $0 < \Delta_K < 1$, 
and the $f_2$ dwarfs are given true separations between 1.0 and 10.0 . 
If the dwarf(s) have $K<14$, they are considered as targets, and if
they have apparent separation $>0.1$ arcseconds they are 
considered to be resolved.  We find that in models where the 
fraction of very-close binaries is very large (i.e., $f_1 \gtrsim 0.4$), then
the WFPC2-resolved binary fraction is a significant underestimate of 
the true binary fraction $f_2$.   We therefore
must consider what constraints exist on the fraction of very close,
near-equal luminosity late-M and L dwarf binaries.

\citet{dahn} have published trigonometric parallaxes of 2MASS
and DENIS-selected L dwarfs.  Of the 17
isolated field L dwarfs in their Table 1, and correcting for the
five resolved companions in the range 1-10 A.U., only one (2M1328+21)
appears to be offset from the L dwarf sequence and may be a close,
unresolved double.  This system was observed but
not resolved by \citet{reidhst}.  These data then suggest that
$f_1 \lesssim 10\%$.
This supported by the scarcity of double-lined binaries
detected in high resolution studies of L dwarfs \citep{basrihires,andreas1}.
Similarly, \citet{reidhires} find two double lined systems amongst a sample
of 39 M6.5-L0.5 dwarfs.
Other results come from studies of the Pleiades cluster.  \citet{pleiades}
estimate that 4 of 34 of Pleiades very-low-mass stars and
brown dwarfs are potential near-equal luminosity binaries with separation
$<27$ A.U. based on their positions in the color-magnitude
diagram. 
On the other hand, the double lined system PPl 15 \citep{ppl15}
shows close systems can exist.  

The balance of evidence therefore suggests that $f_1 = 5 \pm 3\%$. 
It is then
likely that the observed binary percentage of $19\pm4\%$ is an overestimate 
of the true binary fraction ($f_2$) in the range 1.6-15 A.U.
We have generated a series of Monte Carlo models similar to 
the 'toy' model described above, but using the observed
distance distribution of the sample.    By using this observed
distance distribution, we include the most important factor in
resolving binaries with HST.  A more advanced teatment
of the luminosity differences as a function of separation would
be beyond the scope of this paper, and would depend on a better
understanding of the sample selection effects,
the substellar IMF, the L dwarf temperature scale, the
star formation history, and other poorly known factors.
Our best estimate, based on the simple Monte Carlo models,
is that the true value $f_2 = 15\pm5\%$; the decrease due to the
bias in including overluminous double is partially offset by 
the inability to resolve all binaries.   
On the other hand, the total binary fraction ($f_{tot} = f_1 + f_2$)
must be larger than $f_2$ and is likely to be close to 20\% under this
model.  
Potential systematic error is comparable to the random errors
due to sample size. In any case, it appears that a binary fraction 
$f_2 = 15 \pm 5\%$, with $15\% \le f_{tot} \lesssim 25\%$.
If enough L dwarfs within 10 parsecs can be discovered, HST
and/or A.O. imaging may be able to improve the 1-3 A.U. 
constraints.

\begin{figure}
\epsscale{0.8}
\plotone{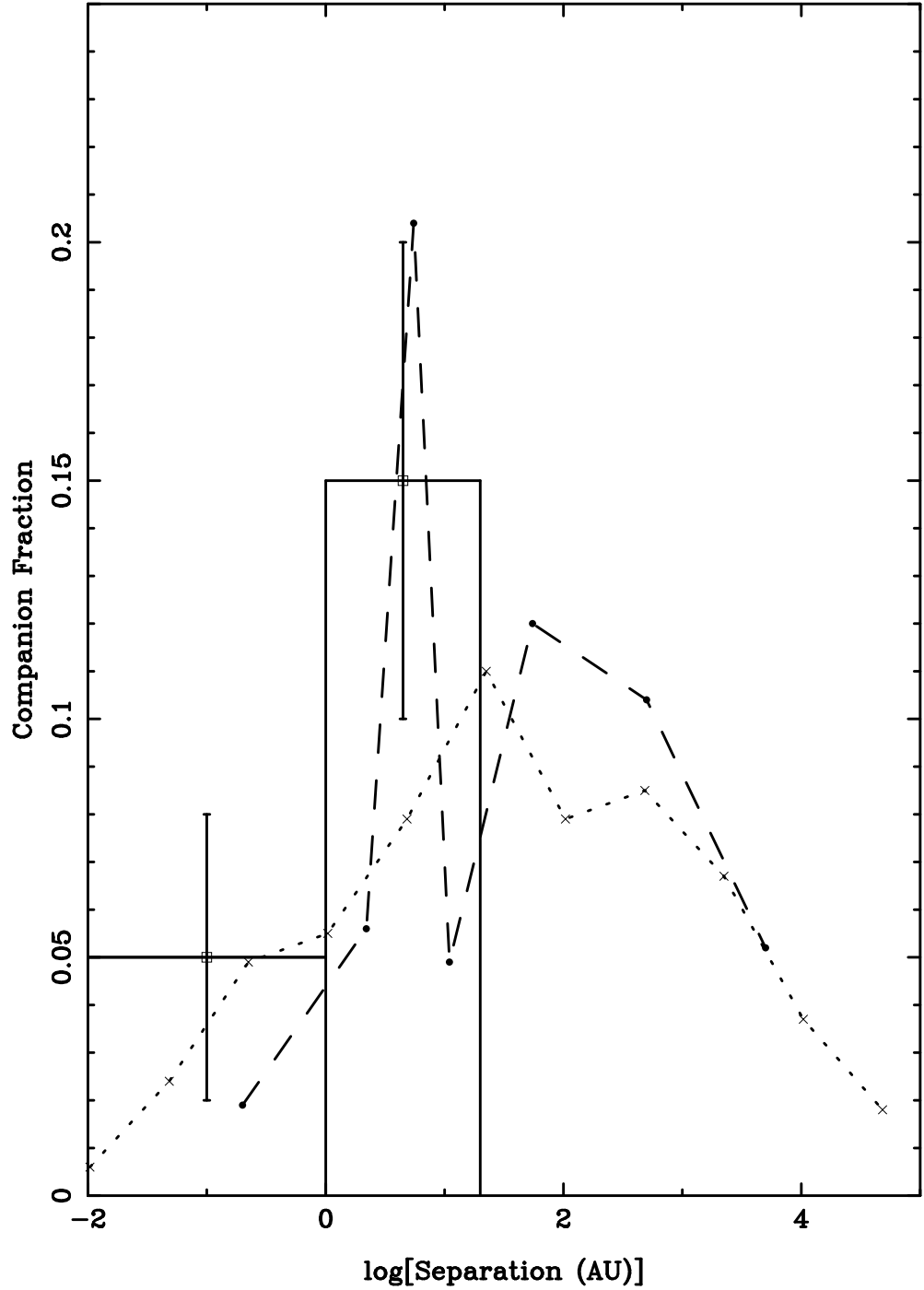}
\caption{Derived binary fractions for late-M and L dwarfs compared to
early-M dwarfs (long-dashed lines, Fischer \& Marcy 1992) and 
G dwarfs (short-dashed lines, Duquennoy \& Mayor 1991); all stellar
companions regardless of mass ratio are included.
Note the lack of wide companions and the relative ``excess'' of
1-16 A.U. companions for the coolest primaries relative to the G dwarfs.  
\label{fig-compare}}
\end{figure}

Since the binary fraction and observed orbital separations of the expanded
sample are similar to those in the \citet{reidhst} sample,
the implications of the results remain the same.  We refer the
reader to \citet{reidhst} and the more recent \citet{close3} 
for a discussion of the similarities and
differences of the M and L dwarf binaries relative to main-sequence
stars.  In particular, the lack of systems at $>20$ A.U. is
now highly significant and a major difference from 
mid-M \citep{fm92} and G dwarfs \citep{dm91}, as seen in 
Figure~\ref{fig-compare}.  A well-defined sample of M5-M6.5 dwarfs
would allow the question of continuity of the binary frequency
and orbital separation distribution to be addressed.

Since we cover both late-M, non-lithium L and lithium L dwarfs, we
may search for differences as a function of mass of the primary.  In other
words, we address the question
``does the binary fraction vary with mass near the hydrogen-burning limit?''
Ten L dwarf systems in this study and \citet{reidhst} have spectroscopic
detections of lithium, placing them below the lithium-burning limit
($\sim 0.055 M_\odot$).\footnote{Three other L dwarf systems, all
apparently single, have marginal lithium detections.}
Three are doubles.  Given the small sample size,
the resulting binary fraction
(30\%) is consistent with that of the non-lithium L dwarfs and
the late-M dwarfs.    There is no evidence of a strong variation in
binary frequency over the mass range $\sim 0.08$ to $\sim 0.04 M_\odot$.
It is interesting to note that all the doubles with companions in the
range 10-16 A.U. have M8-L0 (probably stellar) dwarf primaries, 
while all of the later-L and lithium L (brown) dwarfs are tighter systems.
Additional data are needed to assess the presently marginal
significance of this effect.  Our results are consistent with
\citet{adamhst}'s sample of ten T (brown) dwarfs in both the frequency
and orbital separation.  In particular, the fact that both of the
resolved T dwarfs have separations $<5$ A.U. is consistent with the 
peak seen for the late-M and L dwarfs separations in Figure~\ref{fig-sephist}.

\section{Summary and Conclusions}

We have identified 13 definite and one candidate late-M and
L dwarf binaries in a sample of 82 2MASS and SDSS selected
field dwarfs.  Including the twenty \citet{reidhst} target systems, the 
observed binary fraction is 17$^{+4}_{-3}$\%.  We argue that
accounting for biases and incompleteness leads to an estimated 
binary fraction $15\pm5$\%.  No systems wider than 20 A.U. are
found.  Within the limits of the small sample sizes, we find no 
evidence for variations as a function of mass
in the binary fraction.  Our results are consistent with the
(largely overlapping) late-M sample of \citet{close3} and
the T dwarf sample of \citet{adamhst}.

Additional near-infrared observations of the brown dwarf double 2M1728+3948
are needed.  We interpret the secondary as an early-T dwarf.  Orbital
motions for these systems should be detectable using HST.  This
offers the opportunity to determine masses. 

The greatest need for the future is a better-defined sample of L
dwarf targets, such as a 2MASS-selected sample of bright L dwarfs 
(Cruz et al., in prep.).  The current sample of L dwarfs, although
large, was not selected with consistent
magnitude and color cuts that have well-understood effects.
Once such a sample is available with HST or AO data, the bias and
selection effects can be better understood.  

\acknowledgments

Based on observations made with the NASA/ESA Hubble Space Telescope, 
obtained at the Space Telescope Science Institute,
which is operated by the Association of Universities for Research 
in Astronomy, Inc., under NASA contract NAS 5-26555. These 
observations are associated with proposal 8581.
Support for this work was provided by NASA through grant number 
HST-GO-08581.01-A from the Space Telescope Science Institute, 
which is operated by AURA, Inc., under NASA contract NAS 5-26555
This publication makes use of data products from 2MASS,
which is a joint project of the
University of Massachusetts and IPAC/Caltech, funded by NASA and NSF.

\clearpage

\begin{table}
\dummytable\label{table-phot}
\end{table}

\include{gizis.tab1}

\begin{table}
\dummytable\label{table-binaries}
\end{table}

\include{gizis.tab2}


\end{document}

%% file: gizis.tab1.tex
\begin{deluxetable}{lrrrrrrrrrrrrr}
\rotate
\tiny
\tablewidth{0pc}
\tablenum{1}
\tablecaption{Target Systems}
\tablehead{
\colhead{Name} &
\colhead{$I$} &
\colhead{$Z$} &
\colhead{$J$} &
\colhead{$H$} &
\colhead{$K_s$} &
\colhead{$\sigma_{I}$} &
\colhead{$\sigma_{Z}$} &
\colhead{$\sigma_{J}$} &
\colhead{$\sigma_{H}$} &
\colhead{$\sigma_{K}$} &
\colhead{d (pc)} & 
\colhead{No.} &
\colhead{Ref.} 
}
\startdata
 2MASS J03264225-2102057 & 19.977& 17.422& 16.111& 14.774& 13.885& 0.027& 0.091& 0.100& 0.075& 0.062& 14& 1 & i \\ 
 2MASS J08564793+2235182 & 19.096& 16.888& 15.647& 14.579& 13.924& 0.041& 0.039& 0.061& 0.055& 0.050&35& 1 & h \\ 
 2MASS J00100368+3436099 & 18.231& 16.332& 15.642& 15.080& 14.391& 0.026& 0.059& 0.066& 0.089& 0.073&97& 1 & e \\ 
 2MASS J00283943+1501418 & 20.439& 17.970& 16.477& 15.226& 14.539& 0.042& 0.127& 0.109& 0.086& 0.071&12& 1 & f \\ 
 2MASS J00303013-1450333 & 20.286& 17.752& 16.325& 15.283& 14.492& 0.029& 0.101& 0.115& 0.102& 0.097&11& 1 & f \\ 
 2MASS J00332386-1521309 & 18.718& 16.612& 15.294& 14.225& 13.397& 0.020& 0.058& 0.055& 0.043& 0.040&31& 1 & i \\ 
 2MASS J02085499+2500488 & 17.196& 15.083& 14.015& 13.110& 12.579& 0.017& 0.031& 0.032& 0.035& 0.037&27& 1 & f \\ 
 2MASS J02243670+2537042 & 19.760& 17.722& 16.550& 15.419& 14.670& 0.034& 0.086& 0.107& 0.083& 0.085&83& 1 & f \\ 
 2MASS J03284265+2302051 & 20.434& 17.493& 16.556& 15.547& 14.833& 0.031& 0.111& 0.155& 0.132& 0.105&17& 1 & f \\ 
 2MASS J03350208+2342356 & 14.687&\nodata& 12.259& 11.654& 11.261& 0.021& \nodata& 0.028& 0.033& 0.025&21& 1 & c \\ 
 2MASS J03370359-1758079 & 19.405& 17.019& 15.594& 14.412& 13.588& 0.021& 0.058& 0.065& 0.051& 0.047&13& 1 & f \\ 
 2MASS J03454316+2540233 & 17.244& 14.996& 13.992& 13.170& 12.665& 0.034& 0.020& 0.028& 0.034& 0.029&27.0& 1 &  \\ 
 2MASS J03505737+1818069 & 15.603& 13.733& 12.951& 12.222& 11.763& 0.017& 0.022& 0.031& 0.034& 0.032&28& 1 & c \\ 
 2MASS J03554191+2257016 & 19.545& 17.318& 16.099& 15.023& 14.247& 0.024& 0.064& 0.087& 0.075& 0.066&43& 1 & e \\ 
 2MASS J07533217+2917119 & 18.802& 16.595& 15.485& 14.489& 13.849& 0.022& 0.046& 0.049& 0.047& 0.055&42& 1 & f \\ 
 2MASS J08014056+4628498 & 19.757& 17.356& 16.287& 15.439& 14.540& 0.029& 0.069& 0.140& 0.146& 0.112&45& 1 & f \\ 
 2MASS J08295707+2655099 & 20.774& 18.238& 17.109& 15.723& 14.685& 0.028& 0.113& 0.201& 0.140& 0.081&40& 1 & f \\ 
 2MASS J08320451-0128360 & 17.388& 15.283& 14.127& 13.309& 12.687& 0.023& 0.026& 0.028& 0.027& 0.031&25& 1 & f \\ 
 2MASS J09141884+2238134 & 18.391& 16.280& 15.304& 14.396& 13.898& 0.021& 0.036& 0.047& 0.045& 0.041&56& 1 & e \\ 
 2MASS J09510549+3558021 & 20.971& 18.177& 17.358& 15.894& 14.966& 0.026& 0.143& 0.256& 0.155& 0.101&52& 1 & f \\ 
 2MASS J10170754+1308398 & 17.414& 15.275& 14.126& 13.194& 12.683& 0.040& 0.047& 0.033& 0.034& 0.032&30& 2 & m \\ 
 2MASS J11023375-2359464 & 20.812& 18.398& 17.040& 15.614& 14.794& 0.025& 0.114& 0.193& 0.125& 0.112&29& 1 & f \\ 
 2MASS J11040127+1959217 & 17.981& 15.751& 14.369& 13.492& 12.974& 0.019& 0.028& 0.031& 0.038& 0.038&13& 1 & h \\ 
 2MASS J11083081+6830169 & 16.554& -7.037& 13.139& 12.227& 11.600& 0.027& 0.000& 0.027& 0.024& 0.031&12& 1 & c \\ 
 2MASS J11122567+3548131 & 18.144& 15.990& 14.573& 13.473& 12.694& 0.064& 0.075& 0.043& 0.041& 0.046&16& 1 & f \\ 
 2MASS J11275346+7411076 & 15.812& 13.871& 13.059& 12.367& 11.971& 0.041& 0.032& 0.031& 0.028& 0.029&36& 2 & c \\ 
 2MASS J12391934+2029519 & 17.326& 15.225& 14.465& 13.611& 13.116& 0.018& 0.035& 0.034& 0.038& 0.036&49& 1 & e \\ 
 2MASS J12392727+5515371 & 18.338& 16.009& 14.670& 13.539& 12.743& 0.032& 0.060& 0.033& 0.036& 0.031&19& 2 & f \\ 
 2MASS J13113921+8032219 & 15.458& 13.552& 12.784& 12.116& 11.721& 0.039& 0.037& 0.022& 0.025& 0.024&33& 2 & c \\ 
 2MASS J14032232+3007547 & 15.415& 13.482& 12.691& 12.008& 11.626& 0.024& 0.027& 0.026& 0.028& 0.026&24& 1 & c \\ 
 2MASS J14111735+3936363 & 17.871& 15.691& 14.641& 13.757& 13.244& 0.049& 0.037& 0.039& 0.041& 0.035&33& 1 & f \\ 
 2MASS J14263161+1557012 & 15.823& 13.774& 12.868& 12.182& 11.709& 0.039& 0.040& 0.031& 0.036& 0.033&26& 2 & c \\ 
 2MASS J14304358+2915405 & 17.583& 15.406& 14.279& 13.420& 12.746& 0.041& 0.041& 0.029& 0.033& 0.030&36& 2 & h \\ 
 2MASS J14342644+1940499 & 18.206& 16.249& 15.561& 14.810& 14.389& 0.017& 0.037& 0.063& 0.081& 0.074&92& 1 & e \\ 
 2MASS J14380829+6408363 & 16.064& 13.906& 12.974& 12.160& 11.659& 0.020& 0.021& 0.022& 0.032& 0.025&19& 1 & j \\ 
 2MASS J14385498-1309103 & 19.079& 16.875& 15.528& 14.516& 13.878& 0.022& 0.041& 0.053& 0.049& 0.057&26& 1 & f \\ 
 2MASS J14493784+2355378 & 18.761& 16.574& 15.801& 15.069& 14.337& 0.027& 0.082& 0.079& 0.093& 0.099&98& 2 & f \\ 
 2MASS J14573965+4517167 & 15.999& 13.954& 13.145& 12.412& 11.923& 0.023& 0.029& 0.027& 0.027& 0.025&27& 1 & c \\ 
 2MASS J15065441+1321060 & 16.844& 14.634& 13.414& 12.412& 11.748& 0.017& 0.017& 0.027& 0.031& 0.028&13& 1 & c \\ 
 2MASS J15261405+2043414 & 19.044& 16.656& 15.623& 14.501& 13.918& 0.019& 0.046& 0.059& 0.052& 0.059&36& 1 & f \\ 
 2MASS J15503820+3041038 & 15.444& 13.765& 12.989& 12.411& 11.924& 0.015& 0.016& 0.030& 0.034& 0.032&29& 1 & c \\ 
 2MASS J15510662+6457047 & 15.802& 13.731& 12.870& 12.149& 11.735& 0.016& 0.021& 0.024& 0.029& 0.029&22& 1 & c \\ 
 2MASS J16000548+1708328 & 19.236& 17.030& 16.098& 15.129& 14.669& 0.052& 0.158& 0.097& 0.076& 0.122&107& 2 & f \\ 
 2MASS J16272794+8105075 & 16.030& 13.995& 13.042& 12.332& 11.874& 0.021& 0.027& 0.024& 0.025& 0.027&22& 1 & c \\ 
 2MASS J16351919+4223053 & 15.723& 13.640& 12.886& 12.210& 11.800& 0.021& 0.027& 0.028& 0.032& 0.028&24& 1 & c \\ 
 2MASS J16561885+2835056 & 20.589& 18.106& 16.926& 16.269& 15.194& 0.028& 0.093& 0.000& 0.000& 0.146&37& 1 & f \\ 
 2MASS J17073334+4301304 & 17.133& 15.005& 13.962& 13.205& 12.657& 0.025& 0.022& 0.025& 0.031& 0.038&27& 1 & h \\ 
 2MASS J17102545+2107155 & 18.411& 16.533& 15.867& 15.022& 14.463& 0.023& 0.061& 0.084& 0.090& 0.112&109& 1 & e \\ 
 2MASS J17114573+2232044 & 20.730& \nodata& 17.100& 15.774& 14.694& 0.030& 0.275& 0.187& 0.113& 0.099&44& 1 & f \\ 
 2MASS J17281150+3948593 & 19.684& 16.917& 15.964& 14.781& 13.898& 0.035& 0.086& 0.081& 0.074& 0.052&23& 2 & f \\ 
 2MASS J17434148+2127069 & 19.193& 17.044& 15.795& 14.780& 14.290& 0.024& 0.067& 0.086& 0.066& 0.097&41& 1 & f \\ 
 2MASS J17433487+5844110 & 17.123& 14.762& 14.016& 13.153& 12.669& 0.037& 0.095& 0.025& 0.030& 0.031&30& 1 & h \\ 
 2MASS J18410861+3117279 & 19.746& 17.393& 16.120& 14.970& 14.180& 0.028& 0.056& 0.100& 0.070& 0.084&28& 1 & f \\ 
 2MASS J20543585+1519043 & 19.660& 17.485& 16.205& 15.443& 14.811& 0.031& 0.059& 0.107& 0.128& 0.110&44& 1 & f \\ 
 2MASS J20571538+1715154 & 19.357& 17.268& 16.107& 15.211& 14.567& 0.031& 0.082& 0.110& 0.090& 0.128&63& 1 & f \\ 
 2MASS J21011544+1756586 & 20.677& 18.155& 16.825& 15.792& 15.173& 0.043& 0.188& 0.178& 0.157& 0.194&26& 2 & f \\ 
 2MASS J21402931+1625183 & 15.723& 13.764& 12.943& 12.270& 11.779& 0.029& 0.027& 0.031& 0.034& 0.033&32& 2 & c \\ 
 2MASS J21474365+1431315 & 16.857& 14.705& 13.842& 13.134& 12.652& 0.051& 0.093& 0.036& 0.032& 0.042&40& 2 & c \\ 
 2MASS J21580457-1550098 & 18.559& 16.263& 14.949& 13.916& 13.148& 0.024& 0.060& 0.037& 0.043& 0.040&17& 1 & l \\ 
 2MASS J22062280-2047058 & 15.007& 13.197& 12.381& 11.704& 11.325& 0.031& 0.034& 0.026& 0.023& 0.029&31& 2 & c \\ 
 2MASS J22064500-4217210 & 19.143& 16.913& 15.569& 14.478& 13.595& 0.028& 0.053& 0.068& 0.056& 0.057&25& 1 & f \\ 
 2MASS J22081363+2921215 & 19.588& \nodata& 15.818& 14.825& 14.086& 0.028& 0.427& 0.090& 0.083& 0.083&17& 1 & f \\ 
 2MASS J22244381-0158521 & 17.759& 15.363& 14.052& 12.803& 12.017& 0.016& 0.034& 0.030& 0.029& 0.029&11.4& 1 & f \\ 
 2MASS J22341394+2359559 & 16.247& 14.156& 13.176& 12.353& 11.835& 0.031& 0.022& 0.022& 0.030& 0.034&22& 1 & c \\ 
 2MASS J22425317+2542573 & 18.124& 15.977& 14.795& 13.754& 13.022& 0.025& 0.031& 0.038& 0.036& 0.038&30& 1 & i \\ 
 2MASS J22443167+2043433 & 20.393& 17.678& 16.405& 14.965& 13.932& 0.026& 0.067& 0.128& 0.071& 0.066&11& 1 & a \\ 
 2MASS J23062928-0502285 & 14.004&\nodata& 11.372& 10.718& 10.288& 0.013& \nodata& 0.022& 0.031& 0.027&13& 1 & c \\ 
 2MASS J23094618+1549045 & 17.789& 15.847& 15.005& 14.343& 13.907& 0.034& 0.039& 0.059& 0.058& 0.065&67& 1 & e \\ 
 2MASS J23310161-0406193 & 15.536& 13.722& 12.937& 12.289& 11.930& 0.027& 0.030& 0.027& 0.027& 0.029&28& 2 & c \\ 
 2MASS J23494899+1224386 & 15.320& 13.449& 12.615& 11.952& 11.562& 0.023& 0.022& 0.024& 0.026& 0.030&23& 1 & c \\ 
 SDSS J001911.65+003017.8 & 18.174& 15.991& 14.924& 14.173& 13.588& 0.020& 0.044& 0.036& 0.034& 0.037&37& 1 & d \\ 
 2MASS J03440892+0111251* & 17.872& 15.800& 14.725& 13.890& 13.522& 0.017& 0.034& 0.035& 0.036& 0.047&40& 1 & k \\ 
 SDSSp J104325.10+000148.2 & 19.255& 17.041& 16.096& 15.111& 14.709& 0.027& 0.059& 0.098& 0.082& 0.120&73& 1 & g \\ 
 SDSS J143535.72-004347.0 & 19.747& 17.653& 16.461& 15.672& 15.073& 0.022& 0.088& 0.111& 0.115& 0.141&70& 1 & d \\ 
 SDSS J143517.20-004612.9 & 19.142& 17.202& 16.444& 15.676& 15.268& 0.032& 0.068& 0.104& 0.118& 0.174&136& 1 & d \\ 
 2MASS J15483164-0029414* & 18.230& 16.532& 15.689& 15.123& 14.850& 0.060& 0.035& 0.067& 0.083& 0.125&100& 1 & k \\ 
 SDSS J165329.69+623136.5 & 18.371& 16.280& 15.109& 14.389& 13.864& 0.026& 0.033& 0.054& 0.053& 0.067&39& 1 & d \\ 
 2MASS J17232861+6406230* & 18.961& 17.046& 16.323& 15.644& 15.175& 0.018& 0.051& 0.111& 0.142& 0.191&131& 1 & k \\ 
 2MASS J23355849-0013042* & 18.735& 16.849& 15.969& 15.206& 14.688& 0.058& 0.091& 0.079& 0.089& 0.104&105& 1 & k \\ 
 SDSSp J033035.13-002534.5 & 18.855& 16.520& 15.290& 14.419& 13.829& 0.019& 0.049& 0.045& 0.042& 0.051&23& 1 & b \\ 
 SDSSp J053951.99-005902.0 & 17.575& 15.228& 13.986& 13.065& 12.577& 0.019& 0.026& 0.028& 0.028& 0.030&12& 1 & b \\ 
 2MASS J15154719-0030594* & 16.985& 15.034& 14.181& 13.578& 13.144& 0.016& 0.032& 0.027& 0.028& 0.038&45& 1 & b \\ 
\enddata 
\tablerefs{ 
(a) Dahn et al.(2002) 
(b) Fan et al.(2000) 
(c) Gizis et al.(2000) 
(d) Hawley et al.(2002) 
(e) Kirkpatrick et al.(1999) 
(f) Kirkpatrick et al.(2000) 
(g) Schenider et al.(2002) 
(h) Cruz(priv.comm.)  
(i) Gizis(priv.comm.) 
(j) Kirkpatrick(priv.comm.) 
(k) Knapp(priv.comm.) 
(l) Tinney \& Kirkpatrick(priv.comm.)  
(m) Wilson/Cruz(priv.comm.) 
} 
\end{deluxetable} 

%% file: gizis.tab2.tex
\begin{deluxetable}{lrrrrrrrrrrr}
\rotate
\tablewidth{0pc}
\tablenum{2}
\tablecaption{Doubles}
\tablehead{
\colhead{Name} &
\colhead{$I_A$} &
\colhead{$Z_A$} &
\colhead{$\sigma_{I_A}$} &
\colhead{$\sigma_{Z_A}$} &
\colhead{$I_B$} &
\colhead{$Z_B$} &
\colhead{$\sigma_{I_B}$} &
\colhead{$\sigma_{Z_B}$} &
\colhead{Sep. (``)} &
\colhead{PA} &
\colhead{Sep. (AU)} 
}
\startdata
 2M1017+1308 & 18.080 & 15.905 & 0.035 & 0.043 & 18.260 & 16.165 & 0.034 & 0.045 & 89 & 0.100 &   3 \\ 
 2M1127+7411 & 16.370 & 14.478 & 0.036 & 0.025 & 16.803 & 14.792 & 0.021 & 0.023 & 81 & 0.250 &   9 \\ 
 2M1239+5515 & 19.077 & 16.675 & 0.025 & 0.057 & 19.104 & 16.856 & 0.025 & 0.055 & 7 & 0.213 &   4 \\ 
 2M1311+8032 & 16.105 & 14.223 & 0.033 & 0.031 & 16.328 & 14.393 & 0.029 & 0.024 & 167 & 0.300 &  10 \\ 
 2M1426+1557 & 16.148 & 14.108 & 0.034 & 0.035 & 17.291 & 15.218 & 0.026 & 0.023 & 340 & 0.155 &   4 \\ 
 2M1430+2915 & 18.106 & 15.947 & 0.036 & 0.036 & 18.628 & 16.422 & 0.049 & 0.048 & 326 & 0.084 &   3 \\ 
 2M1449+2355 & 19.115 & 16.984 & 0.018 & 0.080 & 20.151 & 17.829 & 0.027 & 0.113 & 64 & 0.133 &  13 \\ 
 2M1600+1708\tablenotemark{a}& 19.912 & 17.644 & 0.048 & 0.157 & 20.072 & 17.942 & 343 & 0.101 & 0.218 & 0.056 &   6 \\ 
 2M1728+3948 & 20.268 & 17.802 & 0.029 & 0.084 & 20.636 & 17.551 & 0.028 & 0.110 & 27 & 0.130 &   3 \\ 
 2M2101+1756 & 21.226 & 18.579 & 0.038 & 0.187 & 21.681 & 19.381 & 0.045 & 0.343 & 107 & 0.232 &   6 \\ 
 2M2140+1625 & 16.030 & 14.130 & 0.021 & 0.018 & 17.245 & 15.122 & 0.031 & 0.043 & 132 & 0.158 &   5 \\ 
 2M2147+1431 & 17.430 & 15.305 & 0.047 & 0.091 & 17.825 & 15.636 & 0.036 & 0.048 & 329 & 0.323 &  13 \\ 
 2M2206-2047 & 15.721 & 13.940 & 0.024 & 0.027 & 15.799 & 13.959 & 0.024 & 0.024 & 57 & 0.160 &   5 \\ 
 2M2331-0406 & 15.568 & 13.784 & 0.018 & 0.022 & 19.363 & 16.864 & 0.039 & 0.066 & 294 & 0.576 &  16 \\ 
\enddata
\tablenotetext{a}{Candidate binary}
\end{deluxetable}